\documentclass[twocolumn,prl,preprintnumbers,superscriptaddress,showpacs]{revtex4}

\usepackage{graphicx}
\usepackage{amsmath,amsfonts,amssymb,amsxtra}
\usepackage{units}

\makeatletter
\DeclareRobustCommand{\chemical}[1]{%
  {\(\m@th
   \edef\resetfontdimens{\noexpand\)%
       \fontdimen16\textfont2=\the\fontdimen16\textfont2
       \fontdimen17\textfont2=\the\fontdimen17\textfont2\relax}%
   \fontdimen16\textfont2=2.7pt \fontdimen17\textfont2=2.7pt
   \mathrm{#1}%
   \resetfontdimens}}
\DeclareRobustCommand{\bchemical}[1]{%
  {\(\m@th
   \edef\resetfontdimens{\noexpand\)%
       \fontdimen16\textfont2=\the\fontdimen16\textfont2
       \fontdimen17\textfont2=\the\fontdimen17\textfont2\relax}%
   \fontdimen16\textfont2=2.7pt \fontdimen17\textfont2=2.7pt
   \mathbf{#1}%
   \resetfontdimens}}
\makeatother

\newcommand{\lasrnio}{\chemical{La_{2-x}Sr_xNiO_4}}
\newcommand{\lasrcoo}{\chemical{La_{2-x}Sr_xCoO_4}}
\newcommand{\lacacoo}{\chemical{La_{2-x}Ca_xCoO_4}}
\newcommand{\lacuo}{\chemical{La_2CuO_4}}
\newcommand{\lanio}{\chemical{La_2NiO_4}}
\newcommand{\lasrcuo}{\chemical{La_{2-x}Sr_xCuO_4}}
\newcommand{\lacoo}{\chemical{La_2CoO_4}}
\newcommand{\lscopd}{\chemical{La_{1.7}Sr_{0.3}CoO_4}}
\newcommand{\lscopv}{\chemical{La_{1.6}Sr_{0.4}CoO_4}}
\newcommand{\lscopf}{\chemical{La_{1.5}Sr_{0.5}CoO_4}}

\newcommand{\vq}{\chemical{{\bf q}}}

\newcommand{\vQ}{\chemical{{\bf Q}}}

\begin{document}


\title{Magnetic correlations in La$_{2-x}$Sr$_x$CoO$_4$ studied by neutron scattering :
possible evidence for stripe phases}

\author{M. Cwik}%
\author{M. Benomar}%
\author{T. Finger}
\affiliation{%
{II}. Physikalisches Institut, Universit\"at zu K\"oln,
Z\"ulpicher Str. 77, D-50937 K\"oln, Germany}
\author{Y. Sidis}
\affiliation{Laboratoire L\'eon Brillouin, C.E.A./C.N.R.S.,
F-91191 Gif-sur-Yvette CEDEX, France}
\author{D. Senff}
\author{T. Lorenz}%
\affiliation{%
{II}. Physikalisches Institut, Universit\"at zu K\"oln, Z\"ulpicher
Str. 77, D-50937 K\"oln, Germany}
\author{M. Braden}%
\email{braden@ph2.uni-koeln.de}%
\affiliation{%
{II}. Physikalisches Institut, Universit\"at zu K\"oln, Z\"ulpicher
Str. 77, D-50937 K\"oln, Germany}

\date{\today}
\begin{abstract}

Spin correlations in La$_{2-x}$Sr$_x$CoO$_4$ ($0.3 \le x \le
0.6$) have been studied by neutron scattering. The commensurate
antiferromagnetic order of \lacoo \ persists on a very short
range up to a Sr content of x=0.3, whereas small amounts of Sr
suppress commensurate antiferromagnetism in cuprates and in
nickelates. La$_{2-x}$Sr$_x$CoO$_4$ with x$>$0.3 exhibits
incommensurate spin ordering with the modulation closely
following the amount of doping. These incommensurate phases
strongly resemble the stripe phases observed in cuprates and
nickelates, but incommensurate magnetic ordering appears only at
larger Sr content in the cobaltates due to a reduced charge
mobility.

\end{abstract}

\pacs{} \maketitle

The coupled order of charge and magnetic degrees of freedom in
the stripe phases in layered cuprates \cite{1} and nickelates
\cite{2} has attracted strong interest due to its possibly
important role in high-temperature superconductivity. Doping
holes into \lanio \ or \lacuo \ rapidly suppresses the
commensurate antiferromagnetism (AFM) of the parent compounds
resulting in incommensurate ordering. In \lasrnio \ only 12 \% of
Sr drive the system into a stripe phase \cite{2,3}.
Even less charges are necessary to suppress the commensurate AFM
in the cuprates. For concentrations slightly above 2 \%, magnetic
incommensurate superstructure reflections appear, which can be
interpreted in the same stripe picture as that in nickelates
\cite{7}, although alternative explanations have been proposed
\cite{spiral}. For larger Sr-concentration, $x
> 0.055$ this ordering is lost and samples become metallic and
superconducting. Nevertheless, the inelastic magnetic
correlations  in \lasrcuo \ are incommensurate and can be
interpreted in terms of dynamic stripes, since the
incommensurability matches the expected stripe modulation
\cite{8}. Upon co-doping \lasrcuo \ with a rare-earth ion
\cite{1} and in \chemical{La_{2-x}Ba_xCuO_4} \cite{9,10},
however, static stripe ordering has been clearly established.

\begin{figure}
\includegraphics[width=0.95\columnwidth]{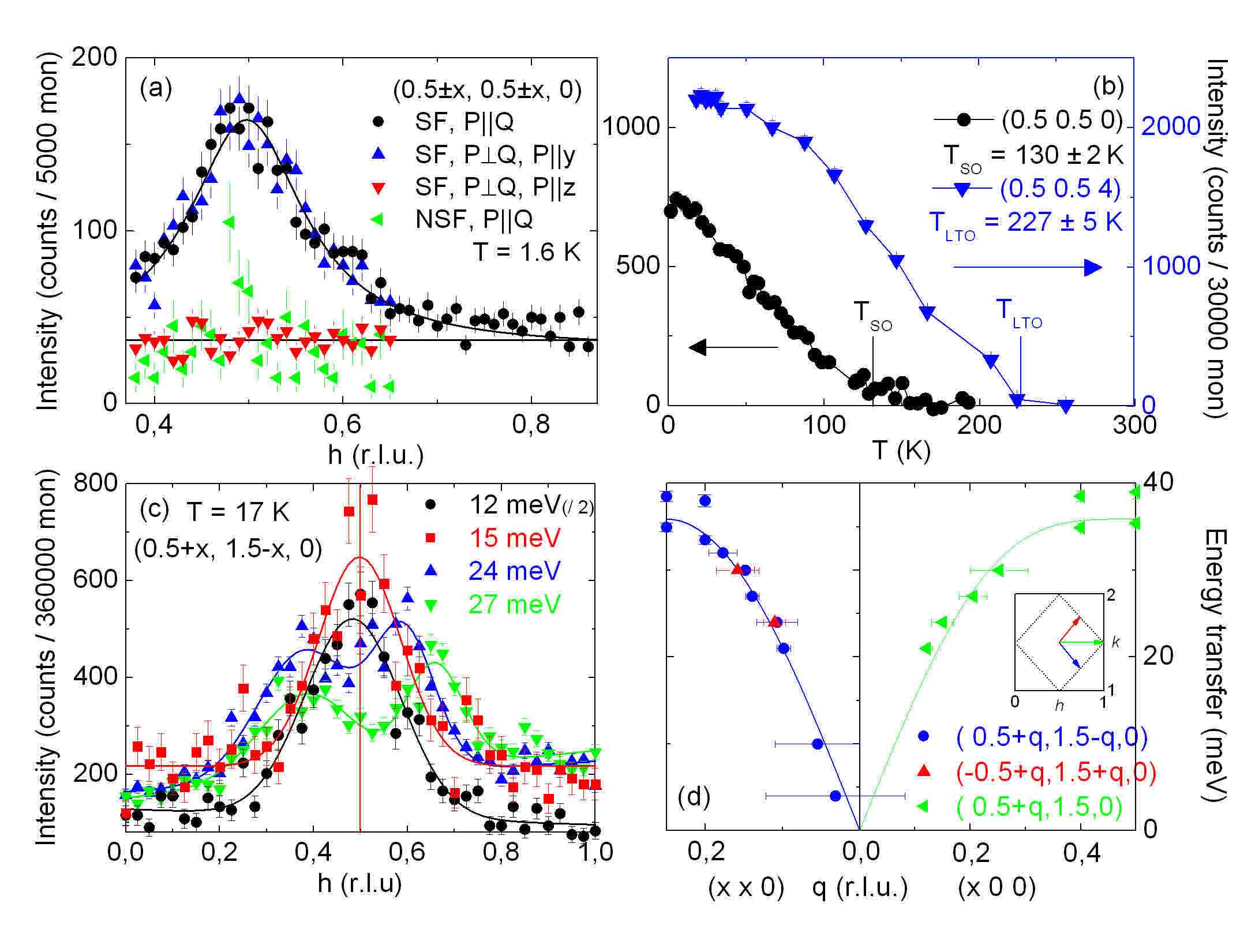}
\vskip -0.4 true cm
\caption{(color online) Magnetic and structural scattering in La$_{1.7}$Sr$_{0.3}$CoO$_4$ . (a) Spin-Flip (SF) and Non-Spin-Flip scattering
(NSF) along ($\frac{1}{2}\pm x$, $\frac{1}{2}\pm x$, 0) for neutron polarization $\mathbf{P}$ perpendicular or parallel $\mathbf{Q}$ at $T=1.6$~K. (b) Temperature
dependence of the magnetic ($\frac{1}{2}$,$\frac{1}{2}$,0) and structural superstructure ($\frac{1}{2}$,$\frac{1}{2}$,4) peak intensities (data in a) and b) were
taken on the 4F spectrometer with polarization analysis and $k_f$=2.66\AA$^{-1}$. c) constant-energy scans across spin-wave excitations at 12\ K. d) Spin-wave
dispersion in La$_{1.7}$Sr$_{0.3}$CoO$_4$ along (x,0,0) and (x,x,0) directions.}
\end{figure}

In analogy with the cuprates and nickelates, it appears
interesting to analyze the possible existence of stripe phases in
\lasrcoo \cite{20}. The parent compound \lacoo \ exhibits
commensurate AFM order ($T_N$=275\ K) similar to \lacuo \ and
\lanio \cite{13}. Furthermore, at half doping, \lscopf ,
checkerboard charge ordering occurs at high temperature,
T$_{CO}$=825\ K, coexisting with magnetic ordering below
T$_N\sim$40\ K \cite{14,15}. The character of the magnetic
ordering between these two compositions, however, has not been
determined so far. We have performed  neutron scattering
experiments on the \lasrcoo -series which reveal an astonishingly
robust commensurate AFM order at low doping and incommensurate
magnetic ordering at intermediate doping in close analogy to the
stripe phases in cuprates and nickelates.

\begin{figure}
\includegraphics[width=0.9\columnwidth]{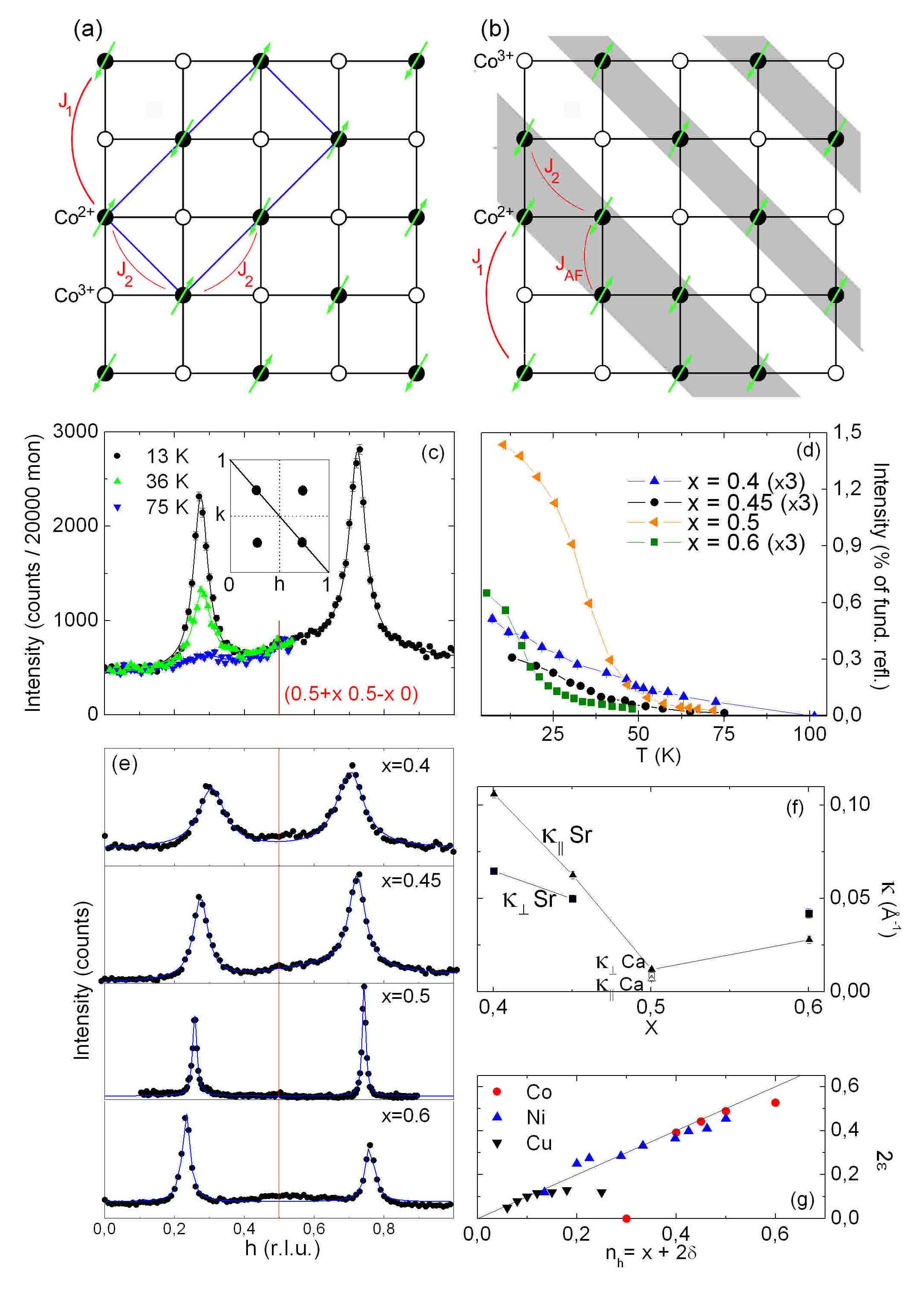}
\vskip -0.5 true cm \caption{(Color online)(a) Magnetic order in the checkerboard charge-ordered phase in \lscopf \ where only Co$^{2+}$ moments contribute; b) the
insertion of an additional Co$^{2+}$ row stabilizes magnetic order through the direct exchange $J_{AF}$. (c) Elastic magnetic scattering in
La$_{1.55}$Sr$_{0.45}$CoO$_4$ for different temperatures (scan direction is shown in the inset); (d) peak heights of the magnetic superstructure and (e) elastic
scans  in \lasrcoo ; (f) in-plane correlation lengths parallel and perpendicular to the modulation for \lasrcoo \ and \lacacoo \; (g) magnetic incommensurability as
a function of Sr-content comparing with \lasrcuo \ \cite{8} and \lasrnio \ \cite{2,3}. }
\end{figure}

\lasrcoo \ and \lacacoo \ single crystals of typically 1cm$^3$
size were grown  in an image furnace following reference
\cite{16}.
The stoichiometry 
was verified by electron-microprobe analysis, by atomic
absorption spectroscopy and by single-crystal as well as by
powder x-ray diffraction \cite{17}. The samples were further
characterized by resistivity and by magnetic susceptibility
measurements indicating for \lscopf \ T$_{CO}$=825(20)\ K and
T$_N$=50(5)\ K, respectively. Elastic and inelastic neutron
scattering experiments were performed using three triple-axis
spectrometers : G4.3, 4F (cold) and 1T (thermal) at the
Laboratoire L\'eon Brillouin.

Pure \lacoo \ 
exhibits a phase transition characterized by a tilt of the CoO$_6$
octahedra \cite{13} leading to a low-temperature orthorhombic
(LTO) phase. Similar to the nickelates and cuprates phase
diagrams, this tilt distortion is rapidly suppressed by the
Sr-doping. In the  \lscopd \ single crystal we find the
characteristic LTO superstructure reflections below T$_{LTO}$=227\
K, see Fig. 1. The longitudinal polarization analysis excludes
any magnetic contribution at this phase transition. However,
below T$_{SO}$=130\ K we find additional superstructure
scattering at (0.5,0.5,$q_l$) whose magnetic origin is proven
through the polarization analysis. The longitudinal polarization
analysis adds an additional selection rule to the general
neutron-scattering law that only magnetic components
perpendicular to the scattering vector \vQ \ contribute: In the
spin-flip channel the magnetic polarization must be perpendicular
to the neutron polarization. The experiment on \lscopd \ was
performed in the [110]/[001] geometry. By measuring the three
spin-flip channels for ${\bf P} || {\bf x}=$(0.5,0.5,0)$=$\vQ ,
${\bf P} || {\bf y}=$(0,0,1)$\bot$\vQ, and ${\bf P} || {\bf
z}=$(1,-1,0)$\bot$\vQ , we may conclude that the ordered moment
fully lies within the $a,b$ plane, see Fig. 1. Magnetic ordering
in \lscopd \ is of the same commensurate nearest-neighbor (nn) AFM
type as that in \lacoo , but the magnetic scattering is very
broad with a Lorentzian width of $\kappa _{ab}$=0.18(2)\
\AA$^{-1}$ \cite{note-sr03}, and there is no detectable
correlation along the $c$ axis. The glass-like nature of the
magnetic ordering is further seen in the magnetic susceptibility
which continuously increases upon cooling through T$_{SO}$ and
which exhibits irreversibility effects only below 16\ K
\cite{hollmann}.

To further characterize the magnetism \lscopd , we have also
anlyzed the magnetic excitations. Typical constant-energy scans
across the AFM zone center are shown in Fig. 1. The small
correlation length together with the steep dispersion prohibit
the separation of low-energy modes, but  spin-wave modes are
found at higher energies confirming that the character of the
magnetic correlation is commensurate AFM. The dispersion is
fitted by spin-wave theory taking only a nn Co$^{2+}$-Co$^{2+}$
interaction into account.
The resulting gap-less spin-wave dispersion,
$\hbar\omega(\mathbf{q})=4J_{AF}S\sqrt{1-\frac{1}{4}\left[\cos{(q_x
2\pi )}+\cos{(q_y 2\pi )}\right]^2}$, describes the observed
spin-wave energies perfectly with $J_{AF}=5.97(8)$~meV and $S=3/2$
indicating an intrinsic Co$^{2+}$-Co$^{2+}$ interaction of the
order of $J_{AF}^{intr.}$=8.5\ meV. Note that here and in the
following the interaction parameters correspond to the energy per
bond, and that the wave-vector $\mathbf{q}$ is given in reduced
units of $\frac{2\pi}{a}$ with $a\sim 3.85$\AA. The \lscopd \
dispersion corresponds to a spin-wave velocity of 138 meV\AA \
from which one may roughly estimate the intrinsic La$_2$CoO$_4$
spin-wave velocity to $\sim$200 meV\AA , lower than values of 340
meV\AA \ in \lanio \ and of 850 meV\AA \ in \lacuo . Energy scans
at the two magnetic zone boundaries in \lscopd \ suggest a weak
splitting, see Fig. 1, which is typical for non-homogeneous
magnets \cite{19}.

The smaller impact of the Sr-doping  on the commensurate
antiferromagnetism in \lasrcoo \ is remarkable in view of the very
strong effects in the nickelates and cuprates, but the impact is
still larger than what is expected for a static non-magnetic
impurity. The substitution of non-magnetic impurities into
layered magnets has been extensively studied \cite{19} for
example in K$_2$(Co$_{1-x}$Mg$_x$)F$_4$. In accordance with
percolation theory long-range AFM order persists up to the
critical concentration of $x_c=0.41$ \cite{18}, whereas the
ordering in \lscopd \ is of short range. In \lasrcoo \ the
magnetic impurity is coupled to the doped charge and may thus
hop. The Co$^{2+}$-sites with 3d$^7$ configuration always stay in
a high-spin (HS) state with $S=3/2$, but at a Co$^{3+}$-site HS,
S=2, intermediate-spin (IS), S=1, and low-spin (LS) states, S=0,
are possible. A Co$^{3+}$ HS state appears unlikely in \lscopd ,
as it should at most weakly perturbate the AFM order. Stronger
effects can be expected for the IS or LS Co$^{3+}$ states where
an efficient trapping of the Co$^{3+}$-site is needed to
stabilize the nn AFM order. Such charge-carrier trapping can arise
from a spin-blockade mechanism as proposed for
HoBaCo$_2$O$_{5.5}$ \cite{21}. In a Co$^{3+}$ LS versus Co$^{2+}$
HS configuration the extra electron at the Co$^{2+}$ site may
only move by passing into the {\it wrong} spin states which
render such processes quite unfavorable.

Let us now turn to the charge and orbital ordering in half-doped
 \lscopf , which has already been studied by Zaliznyak et
al. \cite{14,15}. In our crystal, we find the same
three-dimensional superstructure reflections and perfect
agreement concerning temperature dependencies, T$_N$=48(2)\ K
deduced from the magnetic reflection, and low-temperature
correlation lengths, $\xi _{ab}$=68(3)\ \AA, $\xi _c$=13.1(4)\
\AA \ and $\xi_{ab-charge}$=19(1)\ \AA. The magnetic ordering
does not occur exactly at the commensurate propagation vector of
(0.25,0.25,1) but slightly offset at
$\mathbf{q}$=(0.25$+\delta$,0.25$+\delta$,1) with
$\delta=0.0057(8)$ which is somewhat smaller than the values
observed previously \cite{14,15}. In reference \cite{14} it is
proposed that the charge-ordered arrangement is associated with
non-magnetic LS Co$^{3+}$ sites, whereas an Co$^{3+}$ IS
spin-state is suggested in reference \cite{15}. Our own
structural analysis \cite{17} supports the interpretation of the
LS state corroborated by a quantitative analysis of the
anisotropic magnetic susceptibility \cite{hollmann}. The magnetic
structure with non-magnetic Co$^{3+}$ sites, depicted in Fig. 2a),
perfectly describes the elastic peaks \cite{14} and the full
spin-wave dispersion in La$_{1.5}$Sr$_{0.5}$CoO$_4$ which we have
determined \cite{17} extending a previous study \cite{22}. The
next-nearest neighbor (nnn) Co$^{2+}$-Co$^{2+}$ interaction
($J_1$; linear Co$^{2+}$-O-Co$^{3+}$-O-Co$^{2+}$ path, distance
$2\cdot a$) and the nn Co$^{2+}$-Co$^{2+}$ interaction ($J_{2}$;
distance $\sqrt{2}\cdot a$) are frustrated as  both interactions
are AFM. In mean-field approach the quarter-indexed structure
shown in Fig. 2a) is stabilized for $J_{1} > \frac{1}{2}\cdot
J_2$. Setting $S=\frac{3}{2}$, the magnon dispersion is well
described taking into account only the nnn interaction
$J_1$=2.04(9)meV \cite{17}, which is much lower than $J_{AF}$.
However, $J_1$ couples only half of the Co$^{2+}$ sites in a
single plane (see Fig. 2). Since in addition, $J_{2}$ is almost
fully frustrated \cite{note-frust}, the degenerate in-plane order
is not very stable.

\begin{figure}
\includegraphics[width=0.67\columnwidth,angle=0]{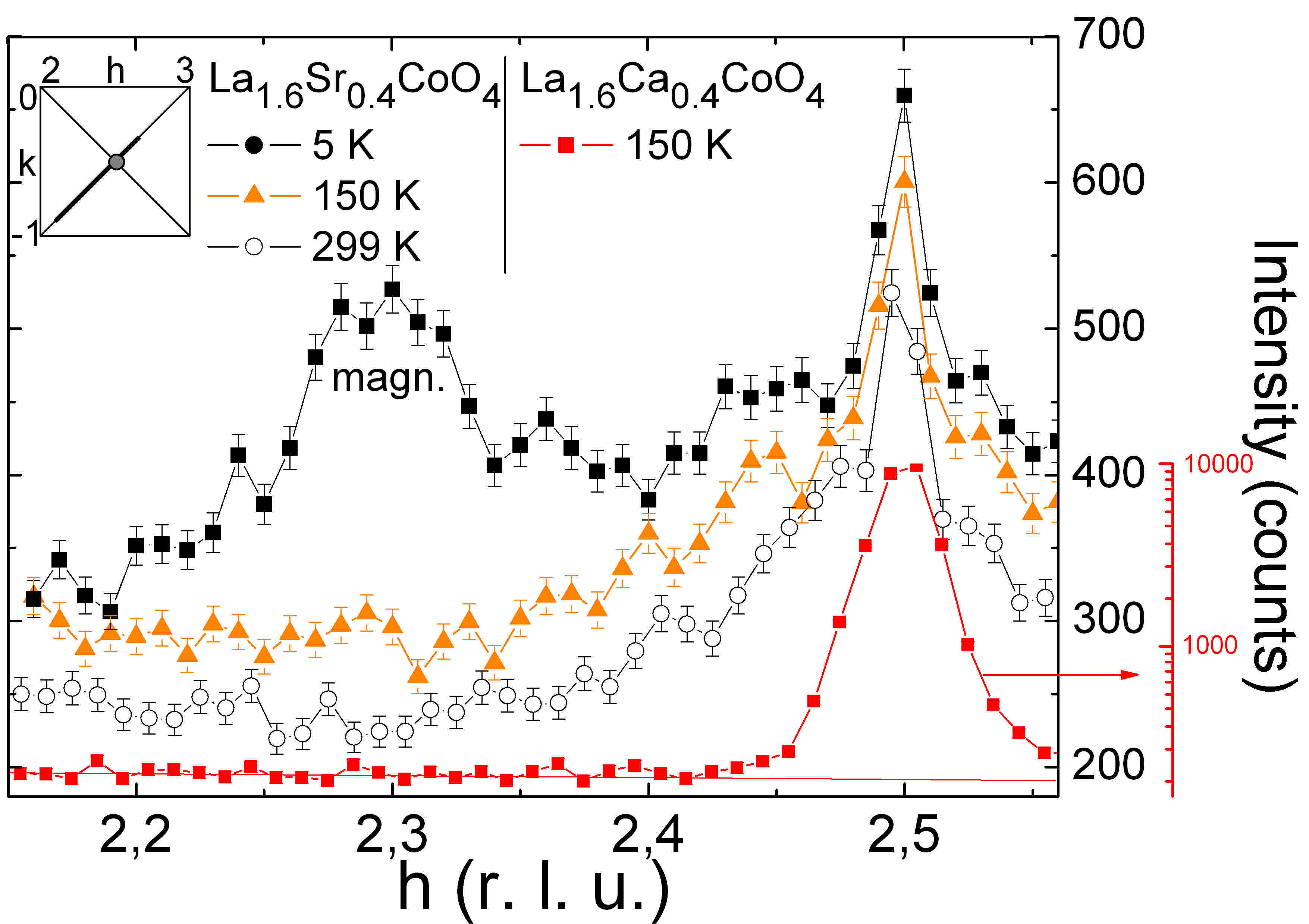}
\caption{Scans across the positions where incommensurate or commensurate scattering related with charge order is expected in \lscopv ~ and in
La$_{1.6}$Ca$_{0.4}$CoO$_4$. }
\end{figure}

Very recently, a HS Co$^{3+}$ moment was proposed for half-doped
La$_{1.5}$Ca$_{0.5}$CoO$_4$ \cite{horigane} basing on the
observation of additional magnetic superstructure reflections at
\vQ =(0.25,0,l) with $l$=$\frac{2n+1}{2}$. We have verified that
these reflections do not appear in our \lscopf \ sample, they
must be at least a factor of 60 smaller than the quarter-indexed
reflections associated with the Co$^{2+}$ ordering therein.

In Fig. 2 we resume the elastic neutron-scattering results for
the intermediate concentrations. Already for x=0.4 there is no
indication for the commensurate AFM ordering; instead,
superstructure reflections arise at ($\frac{1}{2}\pm\epsilon$,
-$\frac{1}{2}\pm\epsilon$, 0) with $2\epsilon$=0.3912(12) which is
very close to the charge carrier content of x=0.4. This magnetic
reflection, thus, perfectly agrees with the diagonal stripe
ordering occurring in the \lasrnio -series. In this picture the
Co$^{3+}$ or Ni$^{3+}$ ions segregate into charged stripes
running along [110] separating AFM stripes, see Fig. 2b). In
consequence the magnetic modulation, $\epsilon$, is determined by
the doped charge concentration: $2\epsilon = x$. Comparable
magnetic superstructure reflections appear in all \lasrcoo \
crystals of intermediate doping, x=0.4, 0.45, 0.5, and 0.6, with
the position following the $2\epsilon =x$ rule. Note, that the
perfect checkerboard ordering in the half-doped compound can be
taken as a stripe phase with $2\epsilon$=0.5 corresponding to an
alternation of Co$^{2+}$ and Co$^{3+}$ rows along the
[110]-direction. The slight incommensurability observed in our
half-doped crystal translates into $2\epsilon$=0.4886(16) only
slightly below the nominal hole content.  The general trend in
the modulation suggests to consider these incommensurate phases
as stripe phases like the analogous phases in nickelates and
cuprates, but, alternatively, the incommensurate magnetic ordering
may be interpreted as a spiral \cite{spiral}, which, however,
leaves the $2\epsilon = x$ relation unexplained.

We have also searched for the corresponding charge-order peaks in
\lscopv \ by scanning diagonally across (2.5,0.5,0), see Fig. 3.
There is sizeable diffuse scattering around \vq =(0.5,0.5,0) in
\lscopv , which is absent in the same scan on a
La$_{1.6}$Ca$_{0.4}$CoO$_4$ crystal of similar size.  Part of the
signal in \lscopv \ can be associated with the superposition of
four broad charge-order peaks at (0.5$\pm$0.1,0.5$\pm$0.1,0) but a
dominant commensurate contribution possibly associated with the
tilt instability prohibits a quantitative analysis.
La$_{1.6}$Ca$_{0.4}$CoO$_4$ exhibits commensurate magnetic order,
and is thus a perfect reference for the background
\cite{note-Ca0.4}.

In Fig. 2(g) we compare the incommensurate modulation vector for
cobaltates, nickelates and cuprates. For \lasrcuo \ we take the
modulation of the inelastic correlation which however directly
reflects that in static stripe phases \cite{8}. The cuprate
modulation for vertical stripes is multiplied by a factor of two
because the hole-occupation in the cuprate stripes amounts only
to one half. The three systems combined follow the ideal linear
relation, but in each of them a saturation of the stripe distance
sets in \cite{3}. In cuprates stripe-like phenomena are observed
already for small hole doping, whereas incommensurate ordering is
shifted to higher doping in nickelates and cobaltates  due to the
reduced mobility of the holes in these materials. In the
cobaltates this effect is by far strongest in agreement with
their much higher electronic resistivity and the spin-blockade
mechanism \cite{21,20}.

\begin{figure}
\includegraphics[width=0.61\columnwidth]{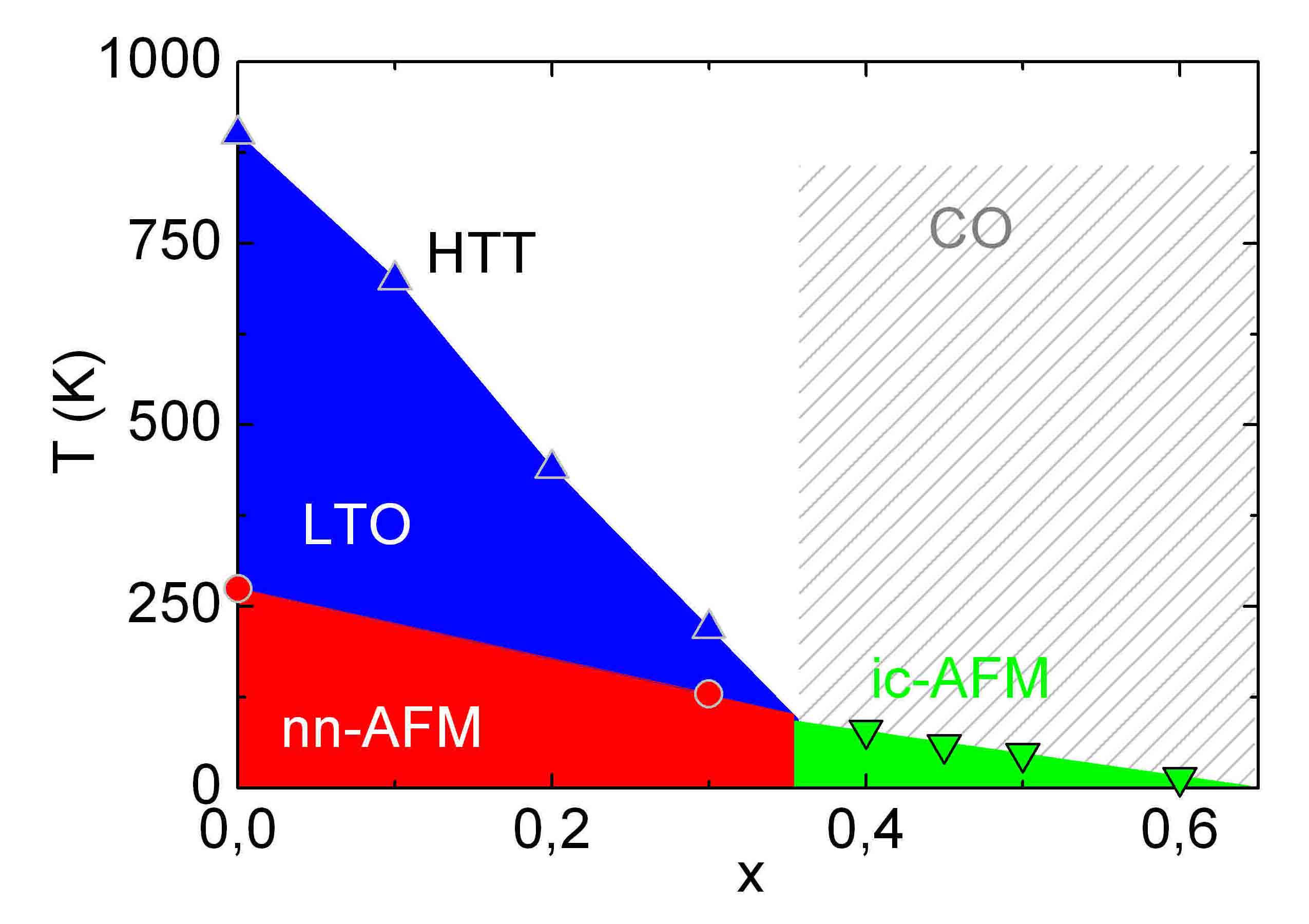}
\caption{ Phase diagram of \lasrcoo . Structural transition temperatures between the tetragonal (HTT) and LTO phases were determined by powder x-ray diffraction
\cite{haider}, $T_N$ of pure La$_2$CoO$_4$ was taken from \cite{13}. Magnetic transition temperatures at finite Sr content were determined extrapolating a linear
fit to the superstructure intensities.}
\end{figure}

Similar to the most stable stripe phases appearing in cuprates and
nickelates at hole-doping levels of x=$1/8$ and $1/3$
\cite{1,3,10}, respectively, there exists a most stable
composition for charge/spin order in the cobaltates as well: It is
the half-doping concentration x=0.5. For this composition we find
the largest in-plane correlation lengths and the strongest
superstructure reflection in comparison to a fundamental
reflection, see Fig. 2. As the correlation lengths for
concentrations away from half doping are reduced, the integrated
magnetic intensity, however, varies much less within the series.
For x=0.5 the magnetic ordering is clearly seen in the magnetic
susceptibility, \cite{22,17}, and the charge order causes an
anomaly in the temperature dependence of the resistitvity
\cite{17}. The fact that, nevertheless, the magnetic ordering is
not fully commensurate (with similar deviations in different
crystals\cite{14,15}) suggests that there is an underlying
intrinsic effect similar to La$_{1.5}$Sr$_{0.5}$NiO$_4$ where the
deviation from commensurability is however six times larger
\cite{23}. A reason for such a deviation might be in both cases
the frustration of the nn interaction $J_2$ and the degeneracy of
the in-plane order mentioned above. The inclusion of additional
magnetic rows lifts the degeneracy and stabilizes magnetic order
due to the strong $J_{AF}$ interaction between neighboring spins,
see Fig. 2b). Also a minor polarization of some ${3+}$ sites may
lift the degeneracy. It is interesting to note, that
La$_{2-x}$Ca$_{x}$CoO$_4$ exhibits commensurate order around half
doping with 2$\epsilon$=0.5016(20) (x=0.4) and
2$\epsilon$=0.5022(18) (x=0.5).

The phase diagram of \lasrcoo , see Fig. 4, qualitatively
resembles those of \lasrnio \ and \lasrcuo \ \cite{3,7,8}. In all
systems the nn AFM order transforms into incommensurate
stripe-like order which is stabilized near a commensurate value.
The magnetic transition temperatures in the cobaltates are
comparable to those found in the co-doped cuprates but
significantly lower than those in nickelates stripe phases, for
example  $T_N$$\sim$150\ K in La$_{1.67}$Sr$_{0.33}$NiO$_4$
\cite{3}. There is a clear trend that the magnetic transition in
\lasrcoo \ continuously decreases with the doping corroborating
the interpretation that the Co$^{3+}$ are magnetically not active.



This work was supported by the Deutsche Forschungsgemeinschaft in
the Sonderforschungsbereich 608. We are thankful to D. Khomskii
for various discussions.


\begin{thebibliography}{28}
\bibitem{note-sr03} This corresponds to a
correlation length of 5.5(5)\AA \ and may suggest a mixture of
ordering schemes, but the high-doping modulation extrapolated to
\vq =(0.65,0.65,0) at x=0.3 can not explain the main signal.
\end{thebibliography}
\end{document}